\documentclass[aps,prl,twocolumn,superscriptaddress,showpacs]{revtex4}

\usepackage{epsfig}
\usepackage{graphicx}
\usepackage{latexsym}

\usepackage{color}

\begin{document}

\title{Two Dimensional Ir-Cluster Lattices on Moir\'{e} of Graphene with Ir(111)}

\author{Alpha N'Diaye}
\email[]{ndiaye@physik.rwth-aachen.de}
\affiliation{I. Physikalisches Institut, RWTH Aachen, 52056 Aachen, Germany}
\author{Sebastian Bleikamp}
\affiliation{I. Physikalisches Institut, RWTH Aachen, 52056 Aachen, Germany}
\author{Peter J. Feibelman}
\affiliation{Sandia National Laboratories, Albuquerque, NM 87185-1413, USA}
\author{Thomas Michely}
\affiliation{I. Physikalisches Institut, RWTH Aachen, 52056 Aachen, Germany}

\date{\today}

\begin{abstract}

Lattices of Ir clusters have been grown by vapor phase deposition on graphene moir\'{e}s on Ir(111). The clusters are highly ordered, spatially and thermally stable below 500\,K. Their narrow size distribution is tunable from 4 to about 130 atoms. A model for cluster binding to the graphene is presented based on scanning tunneling microscopy and density functional theory. The proposed binding mechanism suggests that similar cluster lattices might be grown of materials other than Ir.

\end{abstract}

\pacs{68.65.Cd, 81.16.Dn, 36.40.Sx, 61.46.Bc}

\maketitle

Fabrication of regular arrays of equally sized (monodisperse) clusters on a flat substrate is a central goal for nanotechnology. Owing to their smallness, clusters differ from bulk materials in their chemical and physical properties (cf.~\cite{Meiwes00}). Because these properties depend strongly on size, monodisperse cluster arrays are optimal for fundamental research and applications. Regular arrays of supported clusters are preferable to random ones, because the identical environment of each cluster (e.g. distances from their neighbors) produces a uniform response to external stimulation. Thus, in a regular array, one can use each cluster in the same way, either independently (e.g. for magnetic data storage), or by taking advantage of the coherent collective response of the array as a whole (e.g. in catalysis or for electrical transport). Recent experiments on two-dimensional regular arrays of clusters with a narrow size distribution have explored the size-dependent catalytic activity of Au-clusters \cite{Boyen02}, the magnetic properties of Co-clusters \cite{Weiss05} and electrical transport through PbSe-clusters \cite{Talapin05}, amply demonstrating the usefulness of this approach. 

One route to cluster array fabrication is to deposit atoms or molecules from the vapor phase onto a "template," e.g., a substrate characterized by a periodic array of cluster nucleation sites, to which deposited particles can diffuse. Examples are large unit cell superstructures of oxide films on metal single crystals \cite{Degen04} or regularly spaced steps and surface reconstructions \cite{Weiss05}. Here we demonstrate that graphene moir\'{e}s on an underlying dense-packed metal lattice act as templates for exceptionally well ordered cluster lattices with remarkable properties.

Experiments were performed in an ultra high vacuum, variable temperature scanning tunneling microscopy (STM) apparatus with a base pressure in the $10^{-11}$\,mbar range. Sample cleaning was accomplished by cycles of flash annealing to 1500\,K and sputtering by a mass separated 1.5\,keV Xe$^+$ ion beam at 1100\,K. Ethylene (5\,L) adsorbed at room temperature was thermally decomposed at 1450\,K resulting in the formation of large graphene flakes with sizes around 1000\,\AA~covering about 30\% of the sample surface (compare \cite{Land92}). By continuous exposure of the hot surface to ethylene, samples fully covered by graphene were also prepared. Ir was subsequently evaporated from a current heated Ir-wire, with a standard deposition rate of $3.0 \times 10^{-3}$\,ML/s, where 1 ML is the areal atomic density of the Ir(111) surface. 
Precise coverage calibration was performed by analysis of the fractional area of Ir islands in areas free of graphene.
       
\begin{figure}
\begin{center}
\includegraphics[width=0.95 \linewidth]{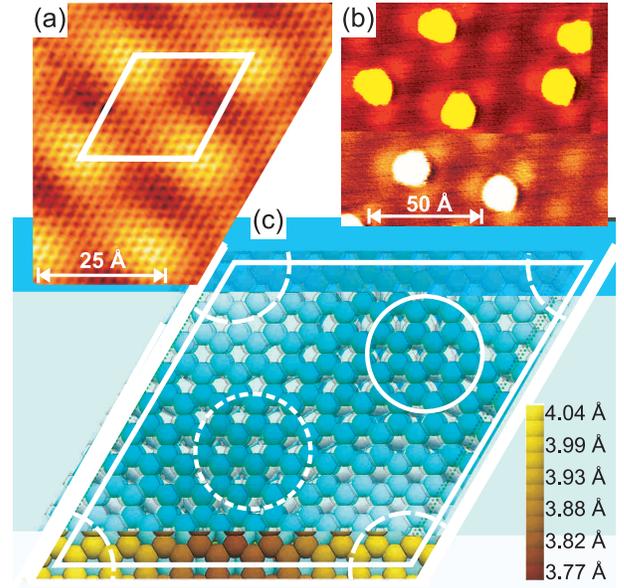}

\caption{(color online)(a) Atomic resolution STM topograph of graphene on Ir(111). The rhombic moir\'{e} unit cell is indicated by lines. Tunneling voltage applied to tip $U_t = +0.2$\,V; tunneling current $I_t = 23$\,nA. (b) STM topograph after deposition of 0.02\,ML Ir on graphene at 350\,K; $U_t = +0.2$\,V; $I_t = 8$\,nA (see text).(c) Schematic illustration of the DFT optimized C(10x10)/Ir(9x9) unit cell. Shading of the C atoms corresponds to their heights as calculated by DFT. 1st, 2nd, and 3rd layer Ir atoms are colored cyan, red and green. Hcp-type region: full circle; fcc-type region: short-dashed circle; atop-type region: dashed circle segments (see text).}

\end{center}
\end{figure}

Fig.\,1(a) is an STM topograph of the graphene moir\'{e} on the underlying Ir(111) substrate lattice. Pronounced bright regions are centered at the corners of its rhombic unit cell, while the darker cell interior shows much less contrast variation. Superimposed on the moir\'{e} is an array of small dark spots with the periodicity of the graphene lattice. Quantitative analysis of the moir\'{e} by LEED and STM yields the following results: The dense-packed $\left\langle  1 0 \bar{1} 0 \right\rangle$-directions of graphene and the unit cell vectors of the moir\'{e} are parallel to the dense-packed $\left\langle 1 \bar{1} 0 \right\rangle$-directions of Ir(111) with angular scatters of $(0 \pm 0.26)^{\circ}$ and $(0 \pm 2.6)^{\circ}$, respectively. The moir\'{e} repeat vectors are of length, $a_{\rm{m}} = (25.3 \pm 0.5)\,\rm{\AA}$ which equals $9.32 \pm 0.15$ times the Ir nearest neighbor distance, $a_{\rm{Ir}}$. The moir\'{e} cell therefore accommodates $A_{\rm{m}} = 87 \pm 3$ Ir surface atoms. 

Fig.\,1(b) displays Ir-clusters grown on graphene at 350\,K by Ir evaporation of 0.02\,ML Ir. Note that the clusters are centered in down-pointing triangles of the moir\'{e}'s bright regions [compare Fig. 1(a)]. Some clusters were removed with the STM tip, prior to imaging, so we could  view the corrugation of the moir\'{e} together with clusters \cite{footnote2}. Owing to tip-surface interaction - often during cluster removal - we also obtained images like Figs.\,1(a) and 1(b) with inverted moir\'{e} contrast. That is, pronounced \emph{dark} regions surrounded the corners of the unit cell, while the brighter cell interior showed only weak brightness variation. In these images, not surprisingly, the clusters are found in down-pointing triangles of \emph{dark} regions.

To shed light on the C-Ir bonding, we optimized the geometry of a thin (3- or 4-layer) Ir(111) slab with a graphene adlayer on its upper surface. The experimental graphene overlayer is not strictly commensurate with Ir(111), but our model supercell, with a $(10 \times 10)$ graphene adlayer on a $(9 \times 9)$ Ir(111) slab is an excellent approximation to reality. For the sake of interpreting the cluster bonding mechanism, we also conducted exploratory calculations with 1, 3 and 4 Ir adatom clusters on the graphene/Ir(111) supercell. 

We performed optimizations using the VASP, Density Functional Theory (DFT) code \cite{vasp1,vasp2} in the PW91 Generalized Gradient Approximation (GGA) \cite{perdew1} with  electron-core interactions represented by the Projector Augmented Wave approximation \cite{bloechl1,kresse1}. The plane-wave cutoff was set to 400\,eV, and Ir-Ir spacings in the rigid one or two lowest slab layers fixed at the PW91 value for bulk Ir, 2.749\,\AA. The surface brillouin zone was sampled by the point, $\bar{\Gamma}$. We accelerated electronic relaxation by Fermi-level smearing (width = 0.2\,eV) \cite{methfessel}, and corrected for the contact potential difference associated with having a graphene adlayer on only one side of the Ir slab \cite{neugebauer}. Systematic DFT error in lattice parameters is not a significant issue for the results. The PW91/GGA calculations imply that a graphene mesh need expand by $< 0.4$\,\% to make a $(10 \times 10)$ graphene cell commensurate with a $(9 \times 9)$ Ir(111) supercell. 

Graphene adsorbed in the experimental angular orientation presents three extremal regions for Ir cluster bonding, fcc-, hcp- and atop-type, named for whether an fcc- or an hcp-hollow, or an Ir atom shows through the local carbon hexagons. [Long-dashed, short-dashed and full circles circumscribe these regions in Fig. \, 1(c).] We infer that Ir clusters bind in hcp-type regions based on the following logic: Fcc- and hcp-type regions are indistinguishable on a 1-layer Ir substrate, whereas atop regions, with all, rather than half the local C-atoms lying in Ir hollows, are structurally and electronically different. Adding a subsurface Ir layer is a second neighbor effect on the adsorbed C atoms, and thus perturbs their bonding weakly. Thus graphene layer properties in fcc- and hcp-type regions remain similar, and the atop-type region quite different. 

Accordingly, we identify the regions with the most pronounced contrast in the STM images [bright regions in Figs.\,1(a) and 1(b)] as atop-type regions. Crystallography then implies that the clusters are adsorbed in the hcp-type and not in the fcc-type regions. Atomically resolved topographs showing graphene and Ir(111) side-by-side support this assignment: Assuming bright protrusions on Ir-terraces to correspond to Ir atoms and dark spots on graphene to centers of carbon atom hexagons, by expanding the Ir lattice registry to the graphene one finds the bright regions in Figs.\,1(a) and 1(b) again to be atop-type ones. It would add weight to our inferences to compare simulated STM images to experiment, but because of moir\'{e} contrast sensitivity to tip condition, we have not.  

Lastly, worthy of mention is that for $T \leq 160$\,K the fcc-type regions of the unit cell also become partly populated with clusters indicating a second, shallower potential energy minimum for Ir on graphene. Cluster adsorption at atop-type areas is never observed. 

Of considerable interest is the nature of the graphene bonds to the metal substrate. Given its strong sp$^2$ bonding, one expects that if graphene is to bind chemically to Ir(111), it must be as a result of C(2p$_{z}$) hybridization with the metal d-bands. Since the 2p$_z$ orbitals point along the normal to the graphene sheet, one expects C atoms in atop sites to form the strongest bonds to the metal through hybridization with Ir(d$_{3z^2-r^2}$) orbitals. The color coded C atom heights above the Ir(111) substrate level in Fig. 1(c) bear out this expectation. The lowest-lying C atoms are in hcp- and fcc-type regions, 3.77\,\AA~and 3.80\,\AA~above an underlying Ir, where many C atoms reside close to atop sites. The highest-lying C atom is found in the atop region 0.27\,\AA~higher than the lowest C of the graphene layer. The markedly different height of the C atoms in the atop-type region is in agreement with the experimental finding that the pronounced moir\'{e} contrast extrema are atop-type regions. 

These results, incidentally, are found insensitive to whether the graphene layer sits atop a 3- or a 4-layer Ir(111) slab, and whether, in the 4-layer case, only the bottom Ir layer is held rigid, or the bottom two. Adding a fourth Ir layer has essentially no effect on the graphene layer's corrugation or binding energy, and changes its height above the Ir surface by no more than ~0.05\,\AA. 

The calculated average binding, 0.20\,eV/C atom, relative to a free graphene sheet and a clean metal slab, seems barely strong enough to indicate chemical bond formation. But this collective result likely masks formation of relatively strong bonds in the hcp- and fcc-type regions of the moir\'{e} cell, compensated by weak binding in atop-type areas. Evidence for this idea is the variation, in the hcp-type regions of Fig. 1, of the sizes of the openings through which red 2nd layer Ir atoms can be seen. That is a manifestation of in-plane shifts of 1st layer Ir atoms, to maximize hybridization with C(2p$_z$) orbitals. 

Structural optimization for Ir clusters has provided some insight into strong cluster bonding to hcp-type and fcc-type regions of the graphene moir\'{e}. There, three out of six C-atoms in a carbon hexagon sit atop a substrate Ir and can form covalent bonds to it through hybridization of C 2p$_z$ orbitals with Ir(d$_{3z^2-r^2}$) orbitals (compare also \cite{Oshima97}). This disturbs the graphene $\pi$-bonds, "activating" the remaining three C-atoms for bonding to the adcluster. Supporting this reasoning, cluster Ir's in the hcp- and fcc-type regions prefer binding not to C-atoms that lie atop Ir's, but to Ir's located over threefold substrate hollows.

\begin{figure}
\begin{center}
\includegraphics[width=0.7 \linewidth]{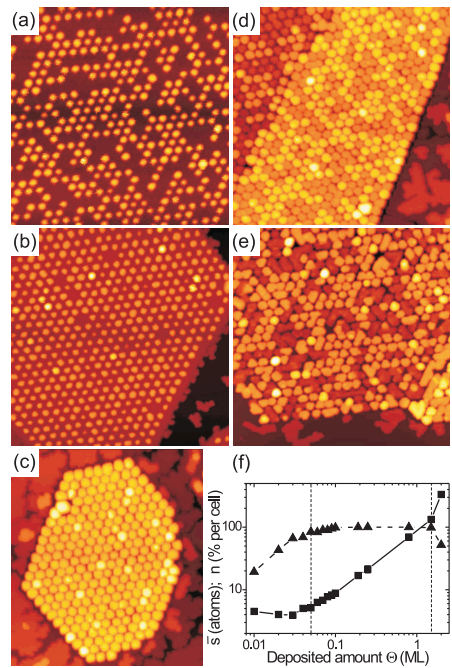}

\caption{(color online) Cluster lattices on graphene flakes. The Ir-coverages are (a) 0.03\,ML, (b) 0.10\,ML, (c) 0.80\,ML, (d) 1.50\,ML and (e) 2.00\,ML. Image height is always 550\,\AA. Unfiltered data. (f) Cluster density $n$  in \% clusters per moir\'{e} cell (full triangles) and average cluster size in atoms $\bar{s}$ versus deposited amount $\Theta$ in ML (full squares). Note that $\bar{s}= \frac{A_{\rm{m}} \Theta}{n}$. Lines to guide the eye.}

\end{center}
\end{figure}

The sequence of STM topographs in Figs.\,2(a)-2(e), taken after Ir deposition at 350\,K together with the quantitative analysis of the cluster density $n$ and the average cluster size in atoms $\bar{s}$ versus deposited amount $\Theta$ in Fig.\,2(f) suggest three regimes of cluster growth. In the nucleation regime represented by Fig.\,2(a) $\bar{s}$ is only weakly dependent on $\Theta$ with $\bar{s} \approx 4-5$, while $n$ increases nearly linearly with $\Theta$. Thus, at least monomers must be mobile on graphene at 350\,K. The absence of monomers and dimers in the cluster size distribution for $\Theta = 0.03$\,ML (not shown) indicates also dimer mobility at 350\,K. The absence of a cluster denuded zone at the edges of the graphene flakes \cite{footnote1} and the high $n$ allow us to conclude that (i) adatoms and dimers are fairly well confined in the unit cell of their arrival or formation and (ii) their intercell mobility is very low on the time scale of deposition (seconds). Therefore jumping of adatoms and dimers to neighboring cells continues after deposition during the $\approx 15$\,minutes of slow cool down from 350\,K to 300\,K prior to STM imaging. After this time, as during imaging no more cluster mobility is observed, no adatoms and dimers are left. In the growth regime for $0.05 \leq \Theta \leq 1.50$\,ML framed by thin vertical dashed lines in Fig.\,2(f) $n$  is nearly [Fig.\,2(b)] or exactly [Fig.\,2(c)] equal to one cluster per moir\'{e} cell and $\bar{s} \propto \Theta$. The growth regime is thus characterized by the slope one in the double logarithmic plot of $\bar{s}$ versus $\Theta$ in Fig.\,2(f). Compared to the Ir-Ir binding the Ir-C binding is weak. Thus, for energetic reasons at some size Ir clusters must become three dimensional. Up to $\bar{s} = 6$ clusters are planar with an apparent height of about 2.2\,\AA~whereas for $\bar{s} \approx 25$ two layer clusters are already as frequent as single layer ones. For $\Theta =0.80$\,ML [Fig.\,2(c)] the average cluster height is 3.1 atomic layers and increases up 4.6 atomic layers for $\Theta = 1.50$\,ML [Fig.\,2(d)]. The observation of four to five layer high uncoalesced clusters with $\bar{s} = 130$ and hexagonal top layer terrace suggests truncated polyhedra composed of the lowest surface energy $\left\{111\right\}$- and $\left\{100\right\}$-facets as typical cluster shape. Clusters in the growth regime are thermally stable up to 500\,K with respect to intercell motion. Assuming an attempt frequency $\nu_0 \approx \frac{k_{\rm{B}}T}{h}$ this translates to a potential energy minimum of a depth $\geq 1.25\,\rm{eV}$. Figure 2(d) with $\bar{s} = 130$ and $\Theta = 1.50$\,ML also marks the transition to the coalescence regime, already exhibiting a few coalesced clusters which extend over two graphene unit cells. It is apparent in Fig.\,2(e) that coalesced clusters extending over several unit cells have a lower height, on average, compared to uncoalesced ones. The coalescence process causes a redistribution of material from upper to lower layers, resulting in a cluster height reduction [the average height in Fig.\,2(e) is 3.7 layers]. Quantitatively the coalescence regime is identified in Fig. 2(f) by a decrease of $n$ and by a superlinear increase of $\bar{s}$.  

Temperature variation alters cluster growth significantly. As mentioned for $T \leq 160$\,K also the fcc-type regions are partly populated, while for $T = 550$\,K and $\Theta = 0.80$\,ML five layer high, triangular clusters grow. These large clusters, with $\bar{s} = 650$ atoms, extend over three hcp-type regions. Although they are still in registry with the moir\'{e}, their positional order is worse than what is found in growth at 350\,K. Extensive temperature dependent measurements are underway to obtain quantitative information on the processes during cluster growth. 

\begin{figure}
\begin{center}
\includegraphics[width=0.55 \linewidth]{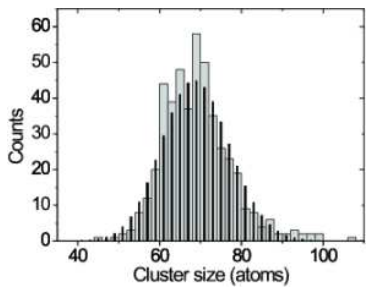}

\caption{Experimental cluster size distribution (gray bars) and corresponding Poisson distribution (narrow black bars) after deposition of 0.80\,ML at 350\,K. Experimental size distribution adjusted to $\bar{s}$ as calculated from $\Theta$. No other corrections for STM-tip effects.}

\end{center}
\end{figure}

A simple model for cluster formation would assume that all atoms deposited into a given unit cell form a single cluster growing in that cell, yielding a Poisson distribution Poi$_{\bar{s}}(s) =\frac{\bar{s}}{s !} e^{- \bar{s}}$ of cluster sizes $s$. As discussed above, this model is inadequate in the nucleation regime for small $\bar{s}$ because of the intercell mobility of adatoms and dimers. However, since adatom and dimer intercell mobility is low \emph{during} the comparatively short deposition time, one expects the experimental distribution to approach Poi$_{\bar{s}}(s)$ as soon as $\Theta$ is large enough that the probability of having deposited fewer than three atoms into a cell is negligible. The experimental distribution is indeed fairly well described by Poi$_{\bar{s}}(s)$ already for $\bar{s}=9$. We note that attachment of adatoms to clusters is irreversible \cite{Busse03} (consistent with the absence of cluster size changes during longtime STM imaging at 500\,K). As for the Poi$_{\bar{s}}(s)$ the standard deviation is $\sigma = \sqrt{\bar{s}}$, the relative standard deviation $\sigma_{\rm{r}} = \frac{\sigma}{\bar{s}} = \frac{1}{\sqrt{\bar{s}}}$ decreases monotonically with $\Theta$. Fig.\,3 exemplifies the agreement of the experimental cluster size distribution with the corresponding Poi$_{70}(s)$ after deposition of 0.80\,ML and demonstrates the narrow size distribution with $\sigma_{\rm{r,exp}} = 12\,\%$.   

In conclusion, when graphene forms a moir\'{e} on a transition metal substrate, with suitable unit cell size, one can  locally functionalize the graphene with adclusters. Monodisperse cluster arrays on the relatively inert graphene surface open new opportunities for catalytic studies. It also might be possible to grow an oxide film on top of a cluster array, which then could be flaked off the graphene, resulting in an array of monodisperse transition metal clusters in an oxide matrix.         

Work by PJF supported by the DOE Office of Basic Energy Sciences, Div. of Mat. Sci. and Eng. Sandia is operated by the Lockheed Martin Company for the USDOE's National Nuclear Security Administration under contract DE-AC04-94AL85000. VASP was developed at the Institut f\"ur Theoretische Physik of the Technische Universit\"at Wien.
Experimental help by Tim Plasa and useful discussions with Norm Bartelt are acknowledged.

\bibliographystyle{unsrt}

\begin{thebibliography}{10}
\bibitem{footnote2}
In order to avoid further cluster removal, Fig. 1(b) had to be taken with an
  increased tunneling resistance compared to Fig. 1(a). Therefore the atomic
  structure of the graphene lattice is hardly visible in Fig. 1(b).
\bibitem{footnote1}
For $\Theta = 0.02 \,{\rm{ML}}$ only the occupation probability of the
  moir\'{e} right at the edge of the flakes is reduced from the terrace value
  of 43\% to 35\%, ruling out any significant transport of adatoms between Ir
  terraces and graphene.
\end{thebibliography}

\end{document}